\documentclass[twocolumn,secnumarabic,amssymb, nobibnotes, aps, prd,superscriptaddress]{revtex4-1}

\usepackage{amsmath}
\usepackage{bookmark}
\usepackage[T1]{fontenc}
\usepackage{graphicx}
\usepackage[caption=false]{subfig}
\usepackage{paralist}
\usepackage{float}
\usepackage{tabularx}
\usepackage{verbatim}
\usepackage{float}

\newtheorem{definition}{Definition}
\begin{document}
\title{Anomaly Detection of Complex Networks Based on Intuitionistic Fuzzy Set Ensemble}

\author{Jin-Fa Wang}
\affiliation{School of Computer Science and Engineering, Northeastern University, Shenyang, 110819, China}
\author{Xiao Liu}
\email[Corresponding author: ]{liu.xiao.xiao.1881@gmail.com}
\affiliation{School of Computer Science and Engineering, Northeastern University, Shenyang, 110819, China}
\affiliation{School of Biological and Biomedical Sciences, Durham University, Durham, DH1 3LE, UK.}
\author{Hai Zhao}
\affiliation{School of Computer Science and Engineering, Northeastern University, Shenyang, 110819, China}
\author{Xing-Chi Chen}
\affiliation{School of Computer Science and Engineering, Northeastern University, Shenyang, 110819, China}
\affiliation{School of Electrical and Data Engineering, University of Technology Sydney, 2007, Australia.}

\thanks{This work was supported by the National Natural Science Foundation of China under NSFC Grants 61671142 and the Fundamental Research Funds for the Central Universities under Grants 02190022117021.}

\begin{abstract}
Ensemble learning for anomaly detection of data structured into complex network has been barely studied due to the inconsistent performance of complex network characteristics and lack of inherent objective function. In this paper, we propose the IFSAD, a new two-phase ensemble method for anomaly detection based on intuitionistic fuzzy set, and applies it to the abnormal behavior detection problem in temporal complex networks. First, it constructs the intuitionistic fuzzy set of single network characteristic which quantifies the degree of membership, non-membership and hesitation of each of network characteristic to the defined linguistic variables so that makes the unuseful or noise characteristics become part of the detection. To build an objective intuitionistic fuzzy relationship, we propose an Gaussian distribution-based membership function which gives a variable hesitation degree. Then, for the fuzzification of multiple network characteristics, the intuitionistic fuzzy weighted geometric operator is adopted to fuse multiple IFSs and to avoid the inconsistent of multiple characteristics. Finally, the score function and precision function are used to sort the fused IFS. Finally we carried out extensive experiments on several complex network datasets for anomaly detection, and the results demonstrate the superiority of our method to state-of-the-art approaches, validating the effectiveness of our method.
\end{abstract}

\maketitle

Since complex network provides a powerful machinery for effectively capturing inter-dependent relationship between study objects\cite{BOCCALETTI20141,Wang2015,Xiao2017,Wei2016}, constructing complex networks from sequentially observed data for anomaly detection has become an effective means\cite{Akoglu2015}. For instance cyber networks, fraud detection, fault detection in medical claims, engineering systems, sensor networks, climate network and many more domains. However one of key challenges is the non-uniform performance of multiple network characteristics in ensemble methods, when each of network characteristic is regarded as the constituent detector alone.

As the advantages that ensemble methods using multiple algorithms or characteristics have better performance than constituent methods alone\cite{Brown2003}, developing effective ensembles for anomaly detection of complex networks has proven to be challenging task\cite{SUN20171,Alelyani2012,Zimek2014}. Existing research works for anomaly ensembles either combine intermediate outcomes(e.g. network characteristic values) from all constituent detectors\cite{JIANG20121093, Krasichkov2015,Rayana2016,YANG60501}, or induce diversity among their detectors to increase the chance that they make independent errors\cite{Hara2104,ZHIJIE20122072,schubert2012,Venkatesh1951}. However, as mentioned in Ref.\cite{Rayana2014}, above methods inevitably combine the inaccurate results(e.g. noise data) and deteriorate the overall detection performance. Thus Rayana et al.\cite{Rayana2016} proposed SELECT method, which automatically and systematically selects the results from the constituent detectors to combine in a fully unsupervised fashion. However, Kavitha et al.\cite{kavitha2011} adopted the Best First Search method to reduce the problem of effective characteristics selection and to remove some unuseful data characteristics before learning in the intrusion detection system.

Apparently some intermediate results present the non-positive correlation with detection goals. In other word, the noise characteristics could suggest that network is normal/abnormal but it is abnormal/normal in fact, and unuseful characteristics could bring non-deterministic for network state\cite{ZHU2142}. Hence it can be inferred that the good characteristics promote the accuracy of abnormal detection, the bad ones strength the certainty of normal state, but the others increase the uncertainty for judging network state. Undoubtedly it become an uncertain theory problem of multi-characteristics and multi-states.

In this paper, the intuitionistic fuzzy set(IFS)\cite{ATANASSOV1986} is adopted to depict the above uncertain problem. Its key idea is that similar networks probably share certain characteristics, for instance the anomaly detection studies based on single network characteristic such as node closeness\cite{Noh2004}, node betweenness\cite{Linton1977}, node degree\cite{Berlingerio2012}, local clustering coefficient\cite{Watts1998}, network diameter\cite{Wang2017}, network entropy\cite{Armstrong2010}, and network assortativity. Different from the Ref.\cite{kavitha2011}, where it is given a non-null hesitation part about the evaluation of study objects to defined the indeterministic behavior, we use the hesitation degree of IFS depicts the useless of unuseful characteristic, and the non-membership degree of IFS denotes the negative correlation of noise characteristic for one certain network state(e.g. normal, fluctuation or abnormal). Meanwhile a new membership function is proposed to resolve the problem of the hesitation index being a fixed value. Furthermore the intuitionistic fuzzy weighted geometric(IFWG) operator\cite{Xu2006} is introduced to fuse multiple network characteristic IFSs into one IFS about network structure to network states. To obtain the detection result, we use score function and precision function to select best IFS which has maximum membership degree to network state. Thus a method of intuitionistic fuzzy set-based anomaly detection (IFSAD) is developed to find the abnormal network structure. We apply our IFSAD to the anomaly detection in temporal complex network datasets\cite{Eagle2006, GARCIA2014,zhang2013}, where IFSAD method utilizes 11 network characteristic metrics. Extensive evaluation on datasets with ground truth shows that IFSAD outperformance the individual detector(i.e. node size-based and diameter-based), SELECT ensemble method.

For a complex network $g = \{N, E, C\}$, the $N$ and $E$ denote the network node set and edge set respectively, and the $C$ is the collection of multiple network characteristic metrics. If it is sampled with a fixed time window $\triangle t$, one sequence of temporal networks $G=\{g(1), g(2), \dots , g(n)\}$ will be obtained by extracting the snapshot topology structure, where the $g(i)$ describes the inter-dependent relationship of research objects at the time ticks $i$. Assume that there are $p$ characteristics metrics $c_i$ for every sampled network $g(i)$. Then in $n$ sampling networks, existing characteristic matrix $C=\{c\}_{p\times n}$ denotes $p$ characteristics series. Every network characteristic will depict the structure characteristic from different perspectives. For instance, the the number of network nodes and the number of network edges describe the network size. When the anomaly behavior occurs, a large number of nodes or edges disappear suddenly in the network. The network diameter denotes the worst communication path length. Under the intentional attack, network diameter will first increases and then decreases quickly\cite{Wang2017}. In this section, we put forward the intuitionistic fuzzification method for single complex network characteristic.

\begin{definition}[Intuitionistic Fuzzy Set, IFS]
$X$ is a finite universal set, such as the network diameter values ($X = C_{i}$). An intuitionistic fuzzy set $A$ in $X$ is an object having the following form.
\begin{equation}
A=\{ <x, \mu_{A}(x), \gamma_{A}(x), \pi_{A}(x)> | x \in X\}
\label{eq:ifs}
\end{equation}
where the $\mu_{A}(x): X \to [0,1]$ defines the degree of membership of intuitionistic fuzzy set $A$, and $\gamma_{A}(x): X \to [0,1]$ defines the degree of non-membership of the element $x \in X$ to set $A$, with the condition $0 \leq \mu_{A}(x) + \gamma_{A}(x) \leq 1$ for all $x$ in $X$. For each IFS in $X$, $\pi_{A}(x)=  1 - \mu_{A} - \gamma_{A}$ is called the hesitation degree(or intuitionistic index) of $x$ to $A$.
\end{definition}

In order to obtain the IFS $A$ of network characteristic, we define the domain of discourse and partition it. First, the domain of discourse $D = [x_{min} - \varepsilon_1, x_{max} + \varepsilon_2]$ is constructed, where $x_{min}$ and $x_{max}$ are the minimum and maximum of set $X$, and $\varepsilon_1$ and $\varepsilon_2$ are proper positive numbers. Second, the intuitionistic fuzzy C-means clustering algorithm(IFCM)\cite{CHAIRA2011} is used to partition the domain of discourse into $m$ clusterings. This is because that compared with the equal interval division, IFCM can classify a collection of objects into homogeneous clusters of objects. Thus we get the clustering center $V = \{v_1, v_2, \dots, v_c\}$ of universe $D$. Then let
\begin{equation}
d_{i} = \left \{
\begin{aligned}
& x_{min} - \varepsilon_1, \quad i = 0 \\
& (v_{i} +  v_{i+1})/2, \quad i = 1, 2, ..., m -1 \\
& x_{max} + \varepsilon_2, \quad i = m \\
\end{aligned}
\right.
\label{eq:d-func}
\end{equation}
As a result, the universe $D$ is divided into $m$ unequal intervals, i.e. $D = \{ [d_0, d_1], [d_1, d_2],\dots, [d_{m-1}, d_m]\}$. Every $x \in X$ should exist $m$ intuitionistic fuzzy sets $A_i= \{ < x, \mu_{A_i}(x), \gamma_{A_i}(x), \pi_{A_i}(x)|x \in D\}$, where the $\mu_{A_i}(x)$ denotes the membership degree of $x$ in $i$th clustering interval $[d_{i-1}, d_i]$ and $\gamma_{A_i}(x)$ is the non-membership degree of that. In this paper $m$ clusterings correspond to $m$ linguistic variables so as to describe the different network states. When $m = 2$, the linguistic variables are usually defined as normal and abnormal. Then if $m = 3$, it could be normal, fluctuate and abnormal. Then one noise value has a lower $\mu$ to linguistic variable "abnormal", but must have a higher $\gamma$ to the variables "normal" or "fluctuate" correspondingly. Moreover the unuseful value will show a bigger hesitation on every IFS. In a word, every characteristic value should be real reflection of network states.

Existed methods of membership and non-membership function usually give the hesitation degree a fixed value, which is not object. So we adopted the Gaussian function(Eq.\eqref{eq:u-func}) which meets below condition: when the distance between the $x$ and the interval center $v$ is lower, the degree of membership $\mu$ is more close to 1.
\begin{equation}
\mu_{A_{i}}(x) = exp(-(x-\psi_{u_i})^2/{2\sigma_{u_i}^2})
\label{eq:u-func}
\end{equation}
Where $i = 1, 2, \dots, m$, and $\psi_{u_i}$ and $\sigma_{u_i}$ are function parameters. Then the following rules are defined so as to resolve above parameters:
\begin{enumerate}
    \item If $x$ is in the middle of an clustering interval, i.e. $x = v_i$, the membership value $\mu_{A_i}(x)$ = 1.
    \item If $x$ is on the boundaries of an clustering interval, i.e. $x = (v_i - v_{i-1} )/2$, let $\pi_{A_i}(x) = \alpha, (0 \leq \alpha \leq 1)$, then $\mu_{A_i}(x) = (1 - \alpha) / 2$.
\end{enumerate}
Based on above rules, the function parameters are resolved:

\begin{align}
\psi_{u_i} & = v_i \label{eq:u-func-psi} \\
\sigma_{u_i}^2 & = -(v_{i -1} + v_{i})^2/(8ln((1-\alpha)/2)) \label{eq:u-func-sigma}
\end{align}
Thus given a value $x$, the membership values for every clustering interval are calculated by equations\eqref{eq:u-func}, \eqref{eq:u-func-psi} and \eqref{eq:u-func-sigma}. However the non-membership function is calculated based on Yager generating function\cite{BURILLO1996305}. The Yager's intuitionistic fuzzy complement is written as following:
\begin{equation}
\gamma_{A_i}(x) = (1 - \mu_{A_i}^\beta(x))^{1/\beta}, \quad \beta > 0
\label{eq:gamma-func}
\end{equation}
When $\mu_{A_i}(x) = 1$, then $\gamma_{A_i}(x) = 0$, and otherwise vice versa. Therefore the IFS(Eq.\eqref{eq:ifs}) becomes:
\begin{equation}
\begin{split}
A= & \{ < x, \mu_{A}(x), (1 - \mu_{A_i}^\beta(x))^{1/\beta}, \\
& 1- \mu_{A_i}(x) - (1 -  \mu_{A_i}^\beta(x))^{1/\beta}> | x \in D\}
\label{eq:ifs-2}
\end{split}
\end{equation}

Single network characteristic, as the individual detector alone, would be used to detect abnormal in some datasets. But a study of Internet sudden change shows that the changes of different characteristics have non-uniform performance in same anomaly events\cite{Ai2013,HU128901}. So an ensemble method eliminating the non-uniform is essential for multiple IFSs reasoning.

For the temporal sequence of one network characteristic $C_i$, we can compute its domain of discourse $D_i$ and IFS $A_i$. Furthermore equation\eqref{eq:ifs-2} can be extended as following for multiple network characteristics:
\begin{equation}
A_{ij}(c)=\{<c,\mu_{A_{ij}}(c),\gamma_{A_{ij}}(c),\pi_{A_{ij}}(c)>| c \in D_i\},
\label{eq:ifs-3}
\end{equation}
where the $i = 1, 2, \dots, p$, $j = 1, 2, \dots, m$, and the $D_i$ is the clustering of one network characteristic $C_i$. The $A_{ij}(c)$ represents the IFS of the $i$th network characteristic $c$ to the $j$th linguist variable. In other word, the $\mu_{A_{ij}}(c)$ is the membership function of $i$th network characteristic value to $j$th clustering interval of network characteristic sequence $D_i$, the $\gamma_{A_{ij}}(c)$ is the non-membership function of $i$th network characteristic value to $j$th clustering interval of the $D_i$, and the $\pi_{A_{ij}}(c)$ is the hesitation degree.

Finally the IFSs $\boldsymbol{A}$ between $p$ network characteristics and $m$ linguist variables are calculated by carrying out the temporal sequence partition and intuitionistic fuzzy set construction on the training set. Hence, the above problem becomes the multi-IFSs reasoning problem. In this paper, intuitionistic fuzzy weighted geometric operator is introduced to fuse the IFSs of $p$ network characteristics to $m$ linguist variables. Let $\boldsymbol{C}' =[c_1\quad c_2 \dots c_p]^{T}$ denotes the characteristic values of a testing network $g'$. Then we define the equation\eqref{eq:ifs-4} to compute the IFS $\boldsymbol{B}$ of the characteristic collection $\boldsymbol{C}'$ to $m$ linguist variables.
\begin{equation}
\begin{aligned}
\boldsymbol{B}=A\odot \boldsymbol{C'}=\left [
\begin{aligned}
& A_{11}(c_1)&\hspace{-0.8em} A_{12}(c_1) &\dots\hspace{-0.8em} & A_{1m}(c_1)\\
& A_{21}(c_2) & \hspace{-0.8em} A_{22}(c_2) & \dots\hspace{-0.8em}  & A_{2m}(c_2)\\
& \quad \vdots &\hspace{-0.8em} \vdots\qquad & \vdots\hspace{-0.8em} & \vdots \qquad \\
& A_{p1}(c_p) &\hspace{-0.8em}  A_{p2}(c_p) & \dots\hspace{-0.8em}  & A_{pm}(c_p)\\
\end{aligned}
\right ]^{T}
\end{aligned}
\label{eq:ifs-4}
\end{equation}
Where the $\boldsymbol{B}_{ij}=A_{ji}(c_j)$ denotes the membership, non-membership and hesitation of the $j$th characteristic value $c_j$ to the $i$ linguistic variable, and then the row vector $\boldsymbol{B}_i=[A_{1i}(c_1)\quad A_{2i}(c_2) \dots A_{pi}(c_p)]$ is the IFSs that current network depicted by $p$ network characteristics is mapped to the $i$th linguistic variable. As a result, the $\boldsymbol{B}$ describes the intuitionistic fuzzy logic relationship between the network characteristics and linguistic variables. In this paper, the linguistic variables represent network states. Here we found that the column vector $\boldsymbol{B}_j=[A_{j1}(c_j)\quad A_{j2}(c_j) \dots A_{jm}(c_j)]^T$, as the individual detector by using single network characteristic, gives the detection result, if we select one linguistic variable that has maximum membership degree as the network state. In order to judgment whether the network is abnormal or not based on the multivariable, the IFWG(Eq.\eqref{eq:ifwg}) is introduced to fuse the IFSs of multiple network characteristics to the linguistic variables.
\begin{equation}
\begin{split}
IFWG_{\boldsymbol{\omega}}(A_1, A_2,...,A_{p}) = A_1^{\omega_1} \oplus A_2^{\omega_2}\oplus...\oplus A_{p}^{\omega_p} \\
= (\prod_{j=1}^{p}\mu_{A_j}^{\omega_j}, 1- \prod_{j=1}^{p}(1 - v_{A_j})^{\omega_j}))
\end{split}
\label{eq:ifwg}
\end{equation}
Where the $\boldsymbol{w} = [w_1\quad w_2\quad \dots\quad w_p)]^{T}$ is the weight vector of $p$ IFSs  that $w_j \in [0,1]$ and $\sum_{j}^{p}{w_j} =1$. Using the IFWG, the network IFSs to each of linguistic variable $\boldsymbol{B}'$ is calculated, as shown in equation\eqref{eq:ifwg-2}. Where the $\mu_{B'_i}(C')$ is the degree of membership of the network state to the $i$th linguistic variable, the $\gamma_{B'_i}(C')$ and $\pi_{B'_i}(C')$ are the non-membership and hesitation of that.
\begin{equation}
\begin{aligned}
\boldsymbol{B'}
& = IFWG_{\omega}(\boldsymbol{B}) \\
&=\left [
\begin{aligned}
&\{<C', \mu(C'), \gamma(C'), \pi(C')> | C' \in D_1 \} \\
&\{<C', \mu(C'), \gamma(C'), \pi(C')> | C' \in D_2 \} \\
&\qquad \vdots  \\
&\{<C', \mu(C'), \gamma(C'), \pi(C')> | C' \in D_m \} \\
\end{aligned}
\right ]
\end{aligned}
\label{eq:ifwg-2}
\end{equation}

Then, we defined a sort method of multiple IFSs based on the score function\cite{CHEN1994163} and precision function\cite{HONG2000103} to obtain the best intuitionistic fuzzy set in the $B'$. For any intuitionistic fuzzy set $A=<\mu, \gamma, \pi>$, the score function $S(A)$ is defined as follows:
\begin{equation}
S(A) = \mu - \gamma, \quad S(A) \in [-1, 1].
\label{eq:s-func}
\end{equation}
It can be seen that the larger the value of $S(A)$ is, the better membership relationship the intuitionistic fuzzy set $A$ is. Then the precision function $H(A)$ of this IFS is defined as follows:
\begin{equation}
H(A) = \mu + \gamma, \quad H(A) \in [0, 1]
\label{eq:h-func}
\end{equation}
It suggests that the larger the value of $H(A)$ is, the higher the precision degree of the intuitionistic fuzzy set $A=<\mu, \gamma>$ is. For any two intuitionistic fuzzy sets $A_1 = <\mu_{A_1}, \gamma_{A_1}, \pi_{A_1}>$ and $A_2 = <\mu_{A_2}, \gamma_{A_2}, \pi_{A_2}>$, the followings hold true:
\begin{enumerate}
\item
  If $S(A_1) < S(A_2)$, then $A_1 < A_2$
\item
  If $S(A_1) = S(A_2)$:
  \begin{enumerate}
\item
  When $H(A_1) = H(A_2)$, then $A_1= A_2$, that is $A_1$ and $A_2$ represent the same information.
\item
  When $H(A_1) < H(A_2)$, then $A_1 < A_2$.
\end{enumerate}
\end{enumerate}
Based on above rules, the $m$ IFSs of the $B'$ can be compared each other. Then the sorted set is given $R=\{B'^{1}, B'^{2}, \dots, B'^{m}\}$, and the linguist variable of $B'^{1}$ mapped is the detection result of our method.

In order to verify the performance of our model, multiple complex networks structured from the Reality Ming-based social datasets\cite{Eagle2006} and botnet-based traffic datasets were used in this paper. We have constructed the sequences of weekly temporal complex networks for three types of interpersonal relationships, i.e. the short messages(SMS), voice call, and bluetooth scans. The ground truth captures semester breaks, exam and sponsor weeks and holidays. Moreover three traffic datasets indicate three botnet scenarios which executed the malicious softwares Rbot and Neris respectively that use several protocols and performed different actions\cite{GARCIA2014}. For the complex networks of botnet traffic, the edge denotes the inter-dependency of network flow within the time influence domain window $\Delta t$ \cite{WANG2016}. The assignment of ground-truth labels is done based on the Ref.\cite{GARCIA2014}. Noted that the Rbot1 and Rbot2 mentioned in the below denote two botnet experimental datasets which executed the Rbot.

We select 11 characteristic metrics of complex network, i.e. node size, edge size, max degree, average degree, k-core, assortativity coefficient, clustering coefficient, structure entropy, shortest path length, network diameter based on maximum, network diameter based on average. Then give the parameters values $\beta = 0.5$, $w_i = 1/p$, and define the linguistic variables $\{normal, fluctuate, abnormal\}$(i.e. $c=3$).

\begin{figure}[ht]
\subfloat[SMS\label{sfig:res-sms}]{%
  \includegraphics[width=.48\columnwidth]{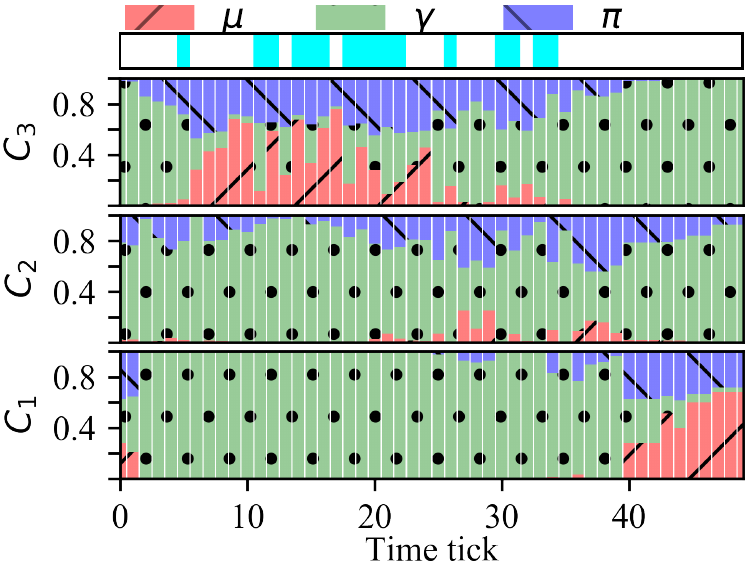}%
}\hfill
\subfloat[Voice\label{sfig:res-voice}]{%
  \includegraphics[width=.48\columnwidth]{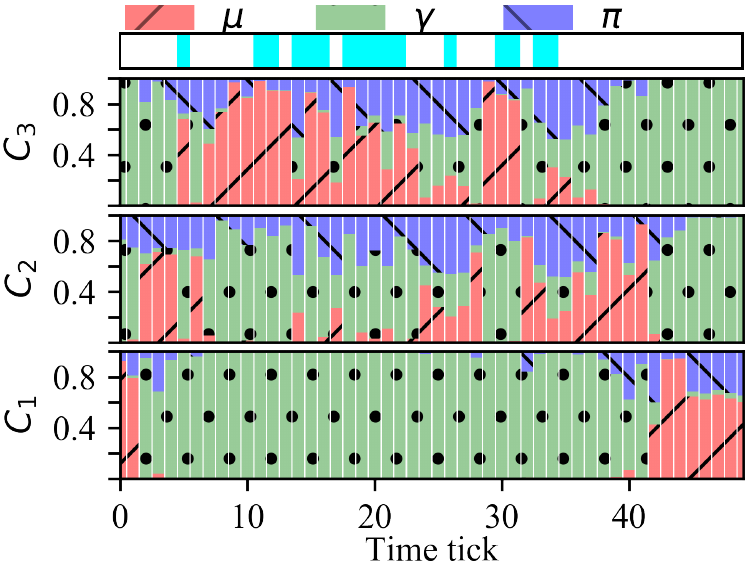}%
}\\
\vspace{-0.4cm}
\subfloat[Bluetooth\label{sfig:res-bluetooth}]{%
  \includegraphics[width=.48\columnwidth]{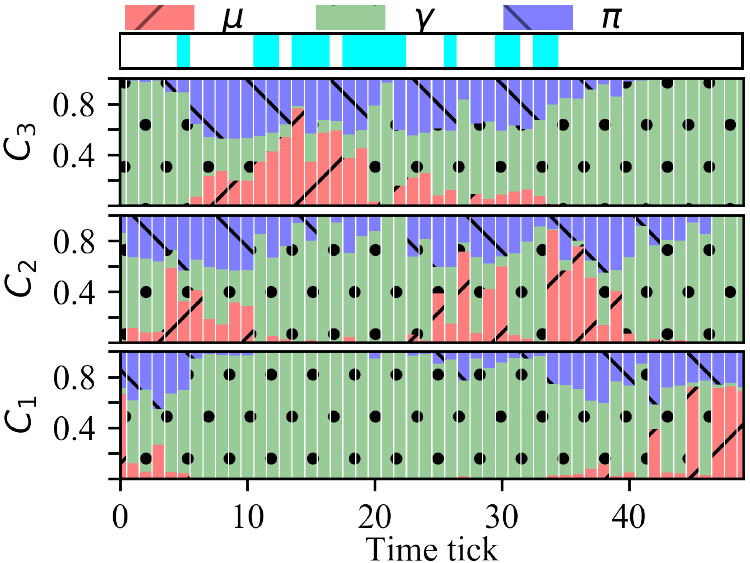}%
} \hfill
\subfloat[Neris\label{sfig:res-bot-9}]{%
  \includegraphics[width=.48\columnwidth]{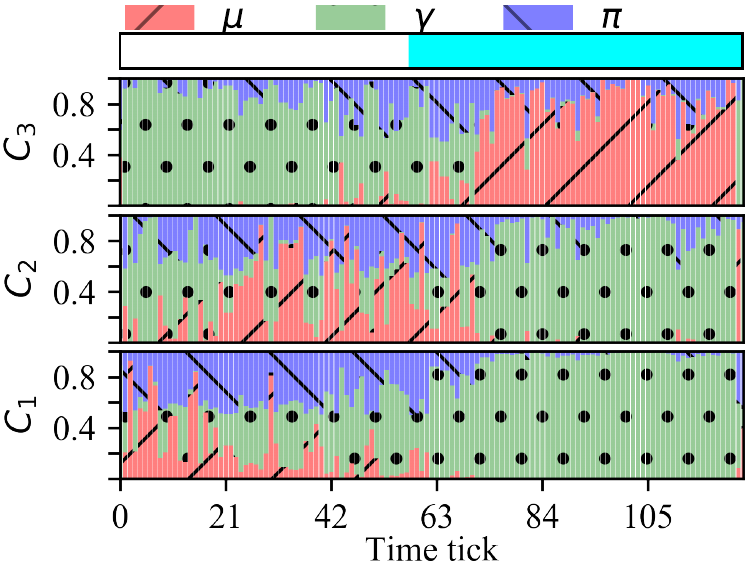}%
}\\
\vspace{-0.4cm}
\subfloat[Rbot1\label{sfig:res-bot-4}]{%
  \includegraphics[width=.48\columnwidth]{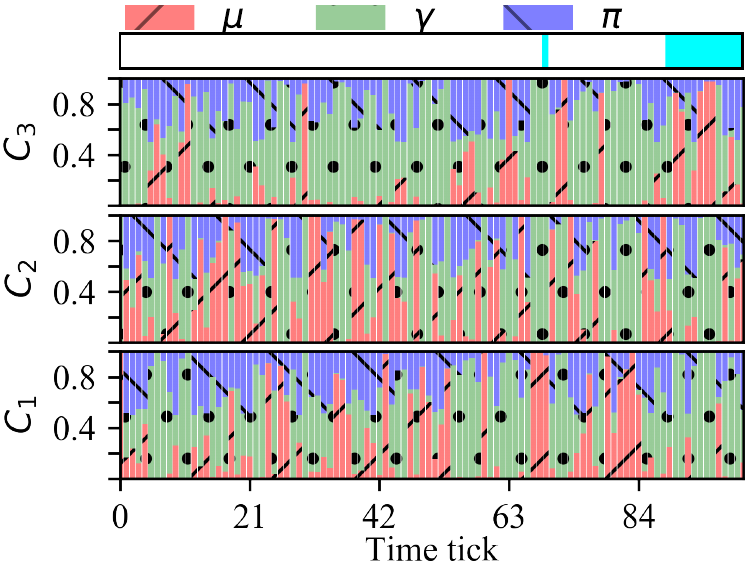}%
}\hfill
\subfloat[Rbot2\label{sfig:res-bot-10}]{%
  \includegraphics[width=.48\columnwidth]{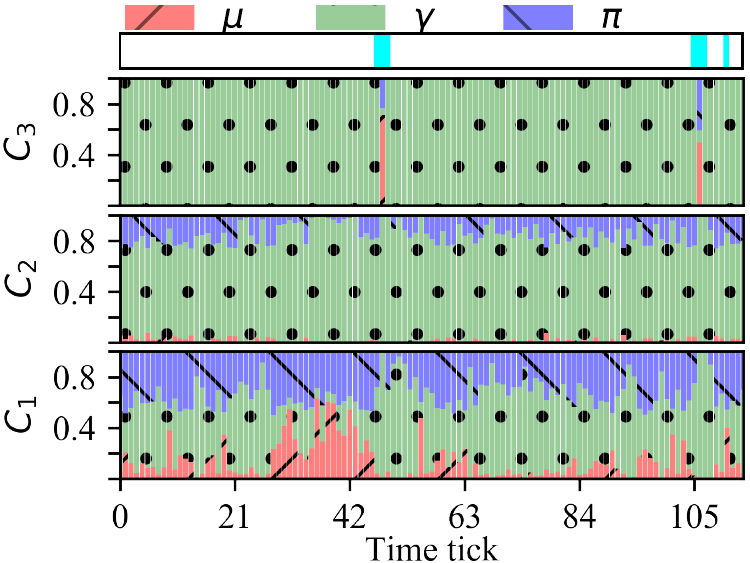}%
}
\vspace{-0.3cm}
\caption{The distribution of membership degree $\mu$ (red histogram with slash), non-membership degree $\gamma$ (green histogram with black point) and hesitation degree $\pi$ (blue histogram with backslash) over three linguistic variables as a function of time ticks. For each of the sequences of temporal network (a)-(f), it is abnormal, if the top ribbon is colored cyan at the time tick $i$, and vice versa. And the three bottom figures depict the abnormal($C_3$), fluctuate($C_2$) and normal($C_1$) state, respectively.}
\label{fig:res-all}
\vspace{-0.5cm}
\end{figure}

First, the effectiveness of our method will be verified in the following. The Fig.\ref{fig:res-all} shows the membership degree $\mu$, non-membership degree $\gamma$ and hesitation degree $\pi$ of sampling networks. Apparently the larger the red bar $\mu$ is, the greater probability the current network state should belong to this linguistic variable. For instance the botnet Neris in Fig.\ref{fig:res-all}(d), the anomaly happens continuously in the second half of the sampled time. At the same time the subfigure $C_3$ shows that the $\mu$ is greater than the $\gamma$ and $\pi$. In the two bottom subfigures $C_2$ and $C_1$, the values of $\gamma$ are in the range of $[0.8, 1]$. It indicates that the state of this network should belong to the linguistic variable "abnormal". Through comparative analysis with the anomaly event, we found that the classification results about network state are very accurate. Moreover based on above analysis method, it can be seen that the others have similarity detection results. Therefore it suggests that our method is effective.

Next the performance of our method is analyzed by using the detection accuracy $a$, detection precision $p$, detection recall $r$ and the $F$-score $F_b$ metrics. The Table \ref{tab:result} shows the results of anomaly detection of our datasets. In this experiment, it was repeated 1000 times, and then calculated the average of each of experimental results so as to decrease the experimental errors. For the detection accuracy $a$, it can be seen that only the SMS network is lower than 90\%, i.e. 84.62\%. The detection accuracy of Rbot2 is best, and the Rbot1 is second best. The common of both is that there are fewer anomaly behaviors in dataset. Comparative analysis between $p$ and $r$, it's inferred that the big number of false positive in the detection results declines the detection precision $p$ of the SMS, Voice and Bluetooth networks, and the big number of false negative in the detection results decline the detection precision $p$ of the Neris, Rbot1 and Rbot2. It inspires us the future work about anomaly detection of complex networks yet. For the values of $F_1$, it suggests that the more abnormal data, the better performance.

\begin{table}[t]
\caption{ The performance of anomaly detection based on our method. The values are the average under 1000 experiments.}\label{tab:result}
\begin{tabularx}{\columnwidth}{ p{0.2\columnwidth} p{0.2\columnwidth} p{0.2\columnwidth} p{0.2\columnwidth} p{0.2\columnwidth}}
\hline
Dataset & $a$ & $p$ & $r$ & $F_1$ \\
\hline
SMS &   0.8462  &   0.7767  &   0.8949  &   0.8134 \\
Voice &   0.9156   &   0.6593  &   0.8557  &   0.7167 \\
Bluetooth  &   0.9243   &   0.8083  &   0.7917  &   0.7840  \\
Neris  &   0.9244    &  0.9832  &   0.8718  &   0.9208\\
Rbot1  &   0.9375   &   1.0000  &   0.6667  &   0.8000\\
Rbot2  &   0.9440   &   0.9000  &   0.4750  &   0.5900\\
\hline
\end{tabularx}
\vspace{-0.8cm}
\end{table}

\begin{figure}[ht]
\begin{center}
\includegraphics[width=.8\columnwidth]{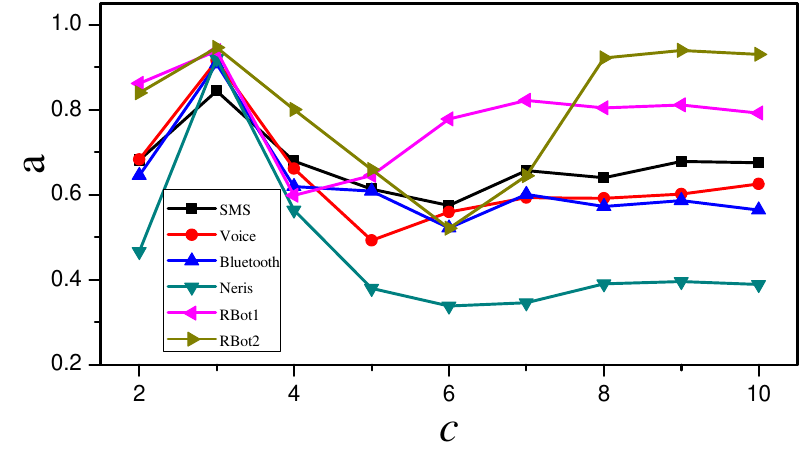}
\vspace{-0.5cm}
\caption{The distribution relationship of the clustering size $c$ as a function of the accuracy $a$ undergoing 1000 experiments}
\label{fig:a-cluster}
\end{center}
\vskip -0.3in
\end{figure}

Then the relationship between linguist variable size $c$ and the detection accuracy $a$ was studied. In the Fig.\ref{fig:a-cluster}, it shows the the trends of SMS, Voice, Bluetooth and Neris networks first increase, then decrease, and finally are stable. However different from the formers in the final stage, the Rbot1 and Rbot2 will increase and then be stable. These results indicates that the $c = 3$ is best for obtaining good detection accuracy.

Finally, the comparison of anomaly detection performance has been implemented by different methods. In this paper, we select three algorithms including network node size\cite{Akoglu2015}, network diameter\cite{gaston2006}, and SELECT\cite{Rayana2016}, where the network node size and network diameter are individual detectors alone by using single network characteristic and SELECT is an ensemble method of combining the outcomes from constituent detectors selectively. However our method make a fuzzification for every characteristic value to linguist variables, and find which represents a best membership degree with multiple characteristics. According to the comparison results in Table\ref{tab:result-compare}, it can be seen that the detection accuracy $a$ of our method are far better than that of node size and network diameter algorithms. Moreover it is also better than the SELECT algorithm except the SMS network.

\begin{table}[t]
\caption{Comparison of the accuracy of anomaly detection by different methods.}\label{tab:result-compare}
\begin{tabularx}{\columnwidth}{ p{0.2\columnwidth} p{0.2\columnwidth} p{0.2\columnwidth} p{0.2\columnwidth} p{0.2\columnwidth}}
\hline
Dataset & Node & Diameter & SELECT & IFSAD  \\
\hline
SMS &    0.7800  &   0.7400  &   0.9217  &   0.8462  \\
Voice   &   0.7600  &   0.6600  &   0.9045  &   0.9156  \\
Bluetooth   &   0.7400  &   0.6600  &   0.8886  &   0.9243  \\
Neris    &   0.6000  &   0.5360  &   0.8781  &   0.9244  \\
Rbot1    &   0.8529  &   0.8529  &   0.8912  &   0.9375  \\
Rbot2    &   0.9478  &   0.9408  &   0.9151  &   0.9440  \\
\hline
\end{tabularx}
\vspace{-0.8cm}
\end{table}

In this work we have proposed IFSAD, a new ensemble method for anomaly detection of structured dataset based on intuitionistic fuzzy set. Our quantitative evaluation for event ensemble on real-word datasets with ground truth show that building the IFS ensembles is effective in boosting detection performance. Overall the anomaly detection performance in computer network traffic is better than that of social network. But the results of the Neris suggests that the more abnormal data, the better performance. All source code of our methods and datasets used in this work are shared openly at \url{http://file.mervin.me/project/cn-ad-ifr}
\\
\\
\\
\\

\end{document}